\documentclass[apjl]{emulateapj}%
\usepackage{amsmath}
\usepackage{graphicx}
\newcommand{\msun}{\rm M_{\sun}}
\newcommand{\Msun}{\rm M_{\sun}}

\newcommand{\erg}{\rm erg}

\newcommand{\s}{\rm s}

\newcommand{\Risco}{R_{\rm ISCO}}

\newcommand{\ka}{{$K\alpha$}}
\newcommand{\spin}{a_{*}}

\shorttitle{Testing the Spin-Jet Correlation}
\shortauthors{Steiner, McClintock, \& Narayan }

\begin{document}

\title{Jet Power and Black Hole Spin: Testing an Empirical Relationship and
  Using it to Predict the Spins of Six Black Holes}

\author{James F.\ Steiner\altaffilmark{1,2}, Jeffrey E.\
  McClintock\altaffilmark{2}, and Ramesh Narayan\altaffilmark{2}}

\altaffiltext{1}{Department of Astronomy, Cambridge University,
  Madingley Road, Cambridge, CB3 0HA, UK.}
\altaffiltext{2}{Harvard-Smithsonian Center for Astrophysics, 60
  Garden Street, Cambridge, MA 02138, USA.}
\email{jsteiner@ast.cam.ac.uk}

\begin{abstract}

  Using 5~GHz radio luminosity at light-curve maximum as a proxy for
  jet power and black-hole spin measurements obtained via the
  continuum-fitting method, \citet{NM12} presented the first direct
  evidence for a relationship between jet power and black hole spin
  for four transient black-hole binaries. We test and confirm their
  empirical relationship using a fifth source, H1743--322, whose spin
  was recently measured. We show that this relationship is consistent
  with Fe-line spin measurements provided that the black hole spin
  axis is assumed to be aligned with the binary angular momentum axis.
  We also show that, during a major outburst of a black hole
  transient, the system reasonably approximates an X-ray standard
  candle. We further show, using the standard synchrotron bubble
  model, that the radio luminosity at light-curve maximum is a good
  proxy for jet kinetic energy. Thus, the observed tight correlation
  between radio power and black hole spin indicates a strong
  underlying link between mechanical jet power and black hole
  spin. Using the fitted correlation between radio power and spin for
  the above five calibration sources, we predict the spins of six
  other black holes in X-ray/radio transient systems with low-mass
  companions. Remarkably, these predicted spins are all relatively
  low, especially when compared to the high measured spins of black
  holes in persistent, wind-fed systems with massive companions.

\end{abstract}


\section{Introduction}\label{section:Intro}

Jets are observed in diverse astrophysical systems and by objects
spanning a wide range of mass: protoplanetary disks around newly
birthed stars, through white dwarfs, neutron stars, stellar-mass black
holes, and up to the supermassive black holes which power active
galactic nuclei \citep{Livio_1999}.  However, despite a wealth of
observational data, the mechanisms responsible for launching and
powering these jets remain uncertain.  In this work, we sharply focus
on one particular class of jets, namely, impulsive ballistic jets
produced during the brightest phase of outbursting black hole
transients.  This is an advantageous approach to the study of jets
because black holes are the simplest astrophysical objects, and also
because, as we will show, the jets we consider are produced at very
nearly the same (Eddington-scaled) mass accretion rates.

In total, there are a few dozen transient X-ray binary systems that
are known to contain black hole primaries \citep{RM06, Ozel_2010}.  A
representative black hole transient is active for about a year and
then quiescent for years or decades before again becoming active.  At
peak flux, a typical system approaches its Eddington
limit\footnote{$L_{\rm Edd}=1.3 \times 10^{39} \erg~\s^{-1} M/10\msun$
  for a black hole of mass $M$.}, and it therefore approximates a
standard candle, as we show in Appendix~\ref{append:candle}.

Based on radio monitoring data collected for several of these X-ray
transients, it is clear that these systems are also radio transients.
Their radio light curves, although of shorter duration, mimic the
X-ray behavior in that they rise rapidly and decay relatively slowly
(e.g., \citealt{Shrader_1994, Brocksopp_2002, Brocksopp_2007}).
However, the peak radio luminosities (unlike the peak X-ray
luminosities) vary widely.  \citealt{NM12} (hereafter NM12) showed for
a sample of four black hole transients that their peak 5~GHz radio
luminosities ranged over a factor of $\approx 250$ while their X-ray
luminosities were all quite similar.  As we show in Section~3, if one
corrects for relativistic beaming, then this range of luminosities is
significantly increased to $\approx 700$ for $\Gamma=2$ and $\approx
1000$ for $\Gamma=5$.

Assuming that the peak radio luminosities of these four transient
sources track the kinetic power of their transient ballistic jets --
an assumption that we show to be reasonable in
Appendix~\ref{append:blob} -- and using the values of their spins
determined via the continuum-fitting method\footnote{A method
pioneered by \citet{Zhang_1997} to measure spin, or to measure mass,
if one assumes a non-spinning black hole \citep{Ebisawa_1991,
Ebisawa_1993}.  The method relies on fitting the thermal disk
component of emission to obtain an estimate of the disk's inner
radius, which is identified with the radius of the innermost stable
circular orbit.  This radius in dimensionless form, $\Risco / M$, is
uniquely and simply related to the black hole's spin
\citep{Bardeen_1972}.  For the mechanics of the continuum-fitting
method, see \citet{McClintock_2006}.}, NM12 reached their central
conclusion: Jet power increases dramatically with increasing black
hole spin $\spin$.  This is the first evidence that jets are powered
by black hole spin, an effect originally predicted by \citet{BZ77}.
NM12 found that jet power scales approximately as $\spin^2$ or
$\Omega_H^2M^2$, where $\Omega_H$ is the angular velocity of the
horizon.  Such a scaling is expected theoretically \citep{BZ77,
Sasha_2010}.

Our result contrasts with an earlier study by \citet{Fender_2010} in
which no correlation was found between jet power and spin.  The
primary difference between the Fender et al. and NM12 studies is the
different proxies used for jet power. Briefly, Fender et al. computed
jet power using a model based on the radio luminosity, X-ray flux and
the rise time of a radio flare event, whereas NM12 simply used the
peak radio luminosity directly as a proxy for jet power.  For a fuller
discussion of the differences between the two studies, we refer the
reader to Section~4 in NM12.

In this paper, we increase from four to five the sample of
microquasars with spins measured via the continuum-fitting method and
with good radio coverage during outburst. Specifically, we add to our
sample H1743--322 whose primary is a slowly spinning black hole, $a_*
= 0.2 \pm 0.3$ \citep{Steiner_2012_H1743}. H1743--322 (hereafter
H1743) is very similar to the microquasar XTE J1550--564 in its X-ray
properties \citep{JEM_H1743} and in its display of pc-scale X-ray and
radio jets \citep{Corbel_2005}. Despite the complete absence of
optical dynamical data -- even the orbital period of H1743 is unknown
-- a kinematic model of the jets allowed a precise determination of
the source distance $D=8.5\pm0.8$~kpc and jet inclination angle 
$i=75\pm3\degr$, which in turn allowed
the spin of this black hole to be measured via the continuum-fitting
method \citep{Steiner_2012_H1743}.

In Section~\ref{section:model}, we present our jet model, and in
Section~\ref{section:h1743}, we use the spin and radio monitoring data
for H1743 to test the NM12 correlation between jet power and spin.  In
Section~\ref{section:predictions}, we first update this correlation by
refitting the data for all five systems, i.e., the four NM12 sources
plus H1743.  Then, as our central objective, we use this correlation
to predict the values of spin for the six black hole primaries in the
following transient systems: GRS~1124-683 (Nova Mus 1991), GX~339--4,
XTE~J1720--318, XTE~J1748-288, XTE~J1859+226 and GS~2000+25.  In
Section~\ref{section:feline}, we compare the correlation based on
continuum-fitting spin data to the available Fe-line spin measurements
for four black hole transients.  Finally, we discuss our results in
Section~\ref{section:discussion} and offer our conclusions in
Section~\ref{section:concs}.

In Appendix~\ref{append:candle}, we validate our ``standard candle''
assumption (NM12) by showing that during major outbursts the systems
we consider reach a substantial fraction of their Eddington limit, and
in Appendix~\ref{append:blob} we describe a simple synchrotron bubble
model and demonstrate that the radio synchrotron flux density at
light-curve maximum is a reasonable proxy for jet kinetic power.

\section{The Jet Power Model}\label{section:model}

We model the bipolar radio jet as a symmetric pair of
isotropically-emitting and optically-thin plasmoids expanding outward
from the core source at a relativistic bulk velocity $\beta$.  The
ratio of observed to emitted flux density for each jet is
\begin{equation}
S_\nu / S_{\nu,0} = \delta^{3-\alpha},
\end{equation}
where $\delta$ is the Doppler factor and $\alpha$ is the radio
spectral index \citep{Mirabel_Rodriguez}.  The Doppler factor of the
brighter (approaching) jet is simply expressed in terms of $\beta$,
the Lorentz factor $\Gamma$, and the jet inclination angle $i$:
\begin{equation}
\delta \equiv \left(\Gamma[1-\beta~{\rm  cos}~i ]\right)^{-1}.
\end{equation}
For the dominant source of emission, i.e.\ the approaching jet, the
observed intensity is greater than the emitted intensity for low
inclinations, and conversely for high inclinations.  For the mildly
relativistic jets of microquasars, $2 \lesssim \Gamma \lesssim 5$
\citep{Fender_2004, Fender_2006}, the Doppler boost becomes less than
unity at intermediate values of inclination in the range $\approx
35-55\degr$.

The NM12 model assumes that the full power of a black hole's ballistic
jet (hereafter, its ``jet power'') is proportional to the peak 5~GHz
radio flux density expressed as a luminosity and scaled by the mass of
the black hole.  The NM12 proxy for jet power is simply
\begin{equation}
P_{\rm jet} = \nu S_{\nu,0}^{\rm tot}  D^2 / M,
\label{eq:pjet}\end{equation}
where $\nu S_{\nu,0}^{\rm tot}$ is the (beaming-corrected) maximum
flux, summed for approaching and receding jets, and $D$ and $M$ are
respectively the distance and mass of the black hole\footnote{In this
  paper, all radio fluxes are referenced to 5~GHz.  None of
  the results here or in NM12 change if we choose a different
  reference radio frequency, e.g., $1.4$\,GHz or $15$\,GHz.  Following
  NM12, we assume a factor of two systematic uncertainty in $P_{\rm
    jet}$.  This is a reasonable error estimate based on the handful
  of available examples of the variations in radio flux observed
  between major outbursts for recurrent transients (see, e.g.,
  \citealt{NM12, MillerJones_2012, Corbel_ATEL1007}).}.  In this work,
jet power throughout has been computed using natural units for these
systems,
\begin{equation}
  P_{\rm jet} = \left(\frac{\nu}{5~{\rm GHz}}\right) \left(\frac{S_{\nu,0}^{\rm tot}}{{\rm Jy}}\right) \left(\frac{D}{{\rm kpc}}\right)^2 \left(\frac{M}{\msun}\right)^{-1}.
\end{equation}
In Appendix~\ref{append:blob}, we show that the approximately linear
relationship between 5~GHz synchrotron emission and bulk kinetic
energy assumed in the empirical NM12 model naturally arises from the
classical synchrotron bubble model for jet ejections.  Any predictions
arising from the use of alternative models or definitions of jet power
are outside the scope of this work.

In the following sections, we compare results obtained using the
empirical NM12 proxy for jet power to the theoretically predicted
scaling between jet power and black hole spin.  The classic work by
\citet{BZ77} describes how spinning black holes interacting with
magnetized accreting gas can act as an engine, tapping into the vast
reservoir of spin energy of the black hole with an efficiency that
depends on magnetic field strength, which in the low spin limit scales
as $P_{\rm jet} \propto \spin^2$. A better approximation, valid over
the full observed range of spins, is
\begin{equation}
P_{\rm jet} \propto (M \Omega_{\rm H})^2, 
\label{eq:pomega}\end{equation}
where $\Omega_H \equiv \spin/(2 M (1+\sqrt{1-\spin^2}))$ is the
angular frequency of the event horizon (for $G=c=1$); it is this
relation that we use throughout.  In recent GRMHD simulations, the
Blandford-Znajek process has been directly demonstrated to have the
capability to efficiently extract black hole spin energy by powering
jets \citep{Sasha_2011}.  We caution that the efficiency of this
process very likely depends on the topology of the magnetic field in
the vicinity of the black hole (e.g., \citealt{Beckwith_2008,
McKinney_Blandford_2009}). Hence the proportionality coefficient in
Equation (\ref{eq:pomega}) is expected to vary with field topology.

\section{Testing the NM12 Correlation}\label{section:h1743}

\subsection{H1743--322}


During major outbursts, the peak radio emission of transient black
hole binaries is associated with powerful X-ray flares.  Such events
are thought to be signatures of jet production and are accompanied by
transitions between hard and soft X-ray states (e.g.,
\citealt{Fender_2004}).  In NM12, our proxy for the jet power of a
source during outburst was computed by simply using the peak radio
flux, which for the four transients considered, as well as many other
transients (e.g., those listed in Table~\ref{tab}), corresponds to the
period of maximum X-ray intensity, with the radio peak usually lagging
shortly behind the X-ray maximum by one or several days.

In the case of H1743, which for months was monitored almost daily in
the X-ray and radio bands \citep{JEM_H1743}, we have additional
information, namely, an accurate estimate of the time of jet ejection,
which occurred on $T_0 =$ MJD 52767.6 $\pm 1.1$ days
\citep{Steiner_2012_H1743}.  Anomalously, the maximum radio flux from
H1743 (96.1 mJy at 4.9 GHz) occurred 30 days prior to the production
of the jets during the early and undistinguished rising phase of the
X-ray source, when its 2--20 keV flux was only 30\% of its maximum
value.  Meanwhile, as we show in \citet{Steiner_2012_H1743}, it was
not until a month later -- at time $T_0$ -- that the jets were
produced by a powerful and impulsive X-ray flare during which the
2--20 keV flux reached an absolute maximum and the 20--200 keV flux
tripled in intensity on a one-day timescale \citep{JEM_H1743}.
Therefore, in computing H1743's jet power, we disregard the maximum
radio flux and instead use the peak radio intensity associated with
the jet launch: $S_{\nu}^{\rm tot} = 34.6$ mJy (4.9 GHz).  We note
that the difference between these two peak flux values is less than a
factor of three; considering the error in spin, both values fall
within $\approx1\sigma$ of the model.  Given also that the peak radio flux in
every other known instance has appeared shortly after the X-ray peak,
we adopt the 34.6 mJy value for H1743.

\begin{figure*}
{\includegraphics[clip=true, angle=90,width=15.85cm]{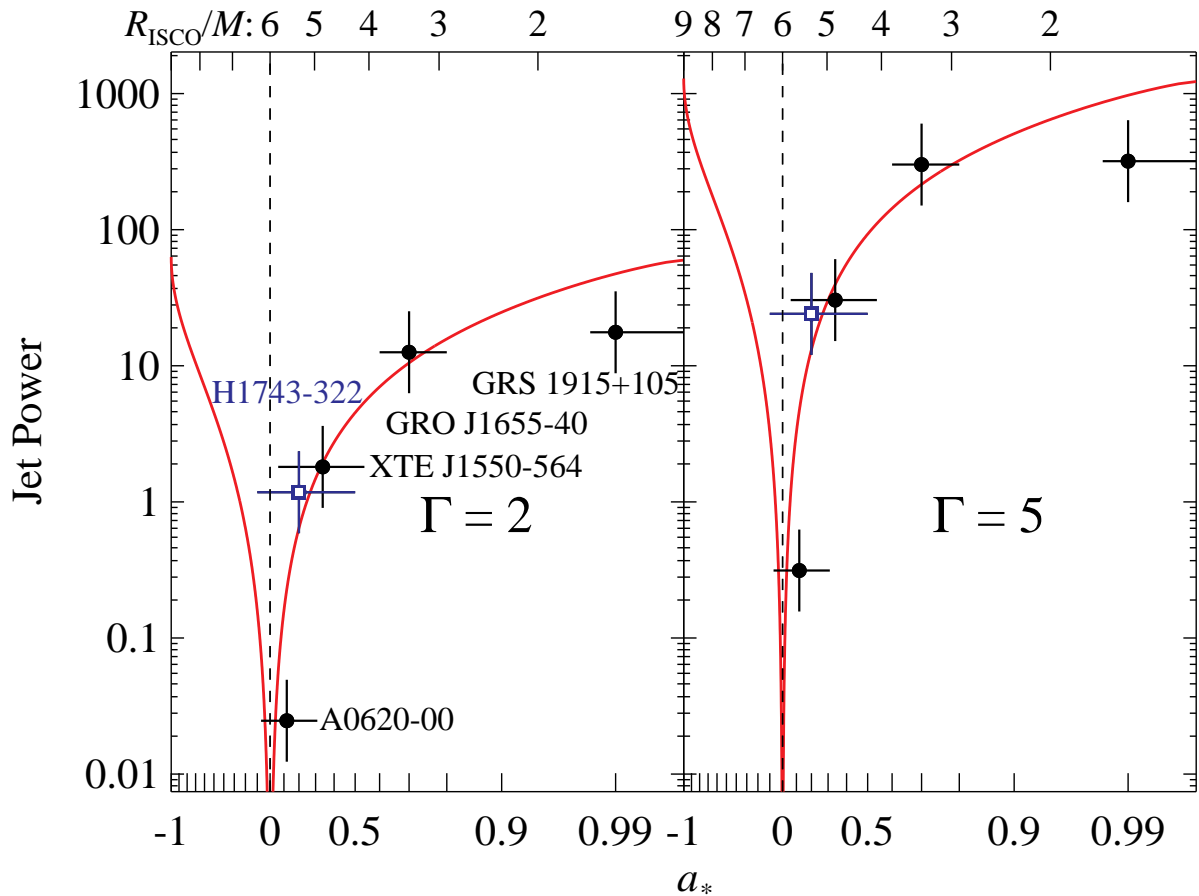}}
\caption{The relationship between radio jet power, and the observable
  $R_{\rm ISCO}/M$ (top axis) and black hole spin (bottom axis).  The
  value $\spin=0$ is marked by a vertical dashed line.  The NM12 data
  are plotted as filled circles and the data for H1743 as an open
  square. The uncertainty in jet power is uniformly assumed to be a
  factor of two.  }\label{fig:h1743}
\end{figure*}

In Figure~\ref{fig:h1743}, the best fit to the NM12 sample of four
black holes is shown\footnote{The best fitting models have
log-normally distributed coefficients:
\begin{eqnarray*}
S_{\nu,0}^{\rm tot} = \left(\frac{\spin}{1+\sqrt{1-\spin^2}}\right)^2 \left(\frac{M}{\msun}\right)\left(\frac{D}{\rm kpc}\right)^{-2}\left(\frac{\nu}{5~{\rm GHz}}\right)^{-1}~{\rm Jy} \\ ~ \qquad ~ \times  \begin{cases}
{\rm Exp(4.2\pm0.5)}, \;\qquad \Gamma = 2 \\
{\rm Exp(7.2\pm0.5)}, \;\qquad \Gamma = 5. 
\end{cases}
\end{eqnarray*}
When H1743 is included, the effects on the curves shown in
Fig.~\ref{fig:h1743} are extremely slight and the changes to the fits
are so small that they are lost within the rounding of the values
given.}, along with the data for H1743 and the four black holes used
to achieve the fit. The data are plotted versus the measurement
quantity $\Risco/M$ (top axis), while the corresponding values of spin
are marked below.  We make the simplifying assumption that $\Gamma$ is
the same for all sources, and we present results using the fiducial
values $\Gamma = 2$ and $\Gamma=5$.  As is evident in the figure, the
data for H1743 are in close agreement with the model of NM12.  We
therefore incorporate H1743 as a fifth calibration source and fit all
five sources to define the relationship between jet power and
$\Omega_{\rm H}$, which we hereafter refer to as the ``NM12 model.''




\subsection{Significance of the Result}

To evaluate the significance of our fitting results (now including
H1743), we have performed a test in which we scrambled the list of
observed fluxes -- with duplicates allowed -- and analyzed these
simulated data sets in the same way that we analyzed the actual data.
We repeated this randomization process 2500 times for each of our two
fiducial values of the Lorentz factor, and in each case a best fit was
obtained.

In less than 1\% of these trials (6/2500 for $\Gamma=2$, 24/2500 for
$\Gamma = 5$) is the fit to the randomized data set as good as the fit
to the actual data.  We conclude that although our sample consists of
only five calibration sources, our empirical correlation is
nevertheless statistically robust.

\section{Predicting the Spins of Black Holes Using the NM12 Model}\label{section:predictions}

We now consider six black holes whose spins have not yet been measured
via the continuum-fitting method.  These systems all displayed a major
X-ray outburst during which the source transitioned through the
thermal dominant (high-soft) state and produced associated radio
flares, which signal the production of jets.  Using data in the
literature, we estimate the jet power of these black holes and thereby
infer their spins by applying the NM12 model.  The names of their host
transient systems, along with estimates of their peak radio fluxes and
distances, are listed in the first three columns of Table~\ref{tab}.
Lower limits on the peak X-ray luminosities achieved by these systems
are given in Table~\ref{tab:appendix}.

In order to estimate jet power, we require estimates of $M$ and the
jet inclination angle $i$, which is problematic for all six systems
listed in Table~\ref{tab}.  XTE J1720-318 and XTE J1748-288 even lack
optical counterparts, while the secondary in GX 339--4 has only been
detected via fluoresced emission lines \citep{Hynes_2003}.  There do
exist literature estimates of $M$ and the orbital inclination angle (a
proxy for $i$) for the remaining four systems.  However, we choose not
to use these estimates of $M$ and $i$ because, in our judgment, the
light-curve and spectroscopic data that are currently available for
these four systems are inadequate to reliably correct the ellipsoidal
light curves for the effects of the strong, variable and
poorly-determined component of disk light (e.g., see
\citealt{Hynes_2005}).  We note that this problem has been largely
overcome for two similar systems, A0620--00 \citep{Cantrell_2010} and
XTE J1550--564 \citep{Orosz_Steiner_2011}, by amassing and analyzing
sufficient data, while it is not a significant problem for other
systems such as GRO J1655--40 \citep{Greene_2001} and 4U 1543--47
\citep{Orosz_1998} because their much more luminous secondaries
strongly outshine their disks.  However, for most black hole
transients, the systematic uncertainties in $M$ and $i$ still remain
sizable \citep{Hynes_2005_coolquiesc, Kreidberg_2012}.

For all six systems listed in Table~\ref{tab}, we adopt the following
approach in assembling the estimates of $M$ and $i$, which we require in
order to estimate jet power.  First, because no firm mass estimates
are available, we use a parametric model for the mass distribution of
black holes in transient systems (Equations A1 \& A2 in
\citealt{Ozel_2012}).

Secondly, we make central use of an eminently
reliable observable, the mass function,
\begin{equation}
f(M) = \frac{M {\rm sin}^3~i}{(1+M_2/M)^2},
\end{equation}
where $M_2$ is the mass of the companion star.  Our results are quite
insensitive to the value of $M_2$ because $M_2/M\ll1$.  In outline,
for the four out of six systems in Table~\ref{tab} with measured
values of the mass function, we use $f(M)$ and the black hole mass
distribution to compute paired values of $M$ and $i$.  We make the
standard assumption that the black hole's spin axis is aligned
perpendicular to the binary orbital plane (see
Section~\ref{section:feline}).  Then, for given values of $\Gamma$ and
$D$ we compute $P_{\rm jet}$.  Finally, we use the NM12 model to infer
the spins of the black holes.

The spin prediction is computed for each black hole using a Monte
Carlo approach as follows.  For 1000 iterations, we consider, with
uniform weighting, a range of jet speeds from $\Gamma = 2$ to $\Gamma
= 5$, and we randomly vary $f(M)$, $D$, and $S_{\nu}$ according to
their measurement errors.  A random value of $M$ is drawn from the
\citet{Ozel_2012} distribution, while $M_2$ is assigned a random value
in the range 0.1--1~$ \msun$.  These six parameters are used to
calculate $i$ and $P_{\rm jet}$.  The data for the five sources of the
NM12 model are then refitted using the selected value of $\Gamma$, and
finally the value of $a_*$ corresponding to $P_{\rm jet}$ is read off
the fitted correlation\footnote{Although both positive and negative
  spin solutions are obtained, we present only the prograde (i.e.,
  $\spin > 0$) result since a retrograde spin has not yet been
  measured (\citealt{McClintock_2011}).  For the adopted model, the
  solutions for each source are symmetric in spin, i.e., the prograde
  and retrograde solutions correspond to the same range of spin apart
  from the sign.}.  Table~\ref{tab} reports the 1$\sigma$ spin ranges
for each source, which are based on the assembled Monte-Carlo results.

  \begin{deluxetable*}{cccccccccccccccccccccccc} 
  \tabletypesize{\scriptsize} 
  \tablecolumns{      9}
  \tablewidth{0pc}  
  \tablecaption{Monte-Carlo Spin Predictions}
  \tablehead{\colhead{Object} & \colhead{$S_{\nu}^{\rm tot}$(5~GHz)~[Jy]} & \colhead{$D$ (kpc)} & \colhead{$f(M)$ ($\Msun$)} & \colhead{Inclination($\degr$)} & \colhead{$\spin$} & \colhead{References}}
  \startdata

GRS 1124--683   & 0.2--1\tablenotemark{a} &  $5.9\pm1.0$    & $3.17 \pm 0.15$ &  44--57   & 0.1--0.4  &  1, 2--7, but see 8 \\  
GX 339--4       &   0.055    &  $8\pm2$        & $5.8 \pm 0.5$                &  54--77   & 0.1--0.4  &  9--11 \\  
                &            &      15        & $5.8 \pm 0.5$                 &  54--77   & 0.2--0.6  &  11, 12 \\  
XTE J1720--318  &  0.0047   &  $6.5\pm3.5$    &   \nodata                     &  \nodata  & $<0.1$    &  13, 14 \\  
XTE J1748--288  &   0.5     &   $8$           & \nodata                       &  \nodata  & 0--0.7  &  15--17 \\    
XTE J1859+226   &   0.10    &  $8\pm3$         & $4.5 \pm 0.6$                &  50--70   & 0.1--0.4  &  4, 18, 19 \\
                &            &    14           & $4.5 \pm 0.6$                &  50--70   & 0.2--0.6  &  19 \\ 
GS 2000+251    & 0.005--0.03\tablenotemark{a} & $2.7\pm0.7$ & $4.97 \pm 0.10$ &  52--74   & $<0.1$  &  1, 6, 20, 21, 22 \\  

\enddata

\tablerefs{(1) \citealt{Jonker_2004};
(2) \citealt{Ball_1995};
(3) \citealt{Gelino_PHD};
(4) \citealt{Hynes_2005};
(5) \citealt{Esin_1997};
(6) \citealt{Barret_1996};
(7) \citealt{Orosz_1996};
(8) \citealt{Shahbaz_1997};
(9) \citealt{Gallo_2004};
(10) \citealt{Zdziarski_2004};
(11) \citealt{Hynes_2003};
(12) \citealt{Hynes_2004};
(13) \citealt{Brocksopp_2005};
(14) \citealt{Chaty_2006};
(15) \citealt{Brocksopp_2007};
(16) \citealt{Mirabel_Rodriguez};
(17) \citealt{Hjellming_BAAS_1998};
(18) \citealt{Brocksopp_2002};
(19) \citealt{Corral_Santana_2011};
(20) \citealt{Hjellming_1988};
(21) \citealt{Filippenko_1995};
(22) \citealt{Callanan_1996}.}

\tablenotetext{a}{The lower limit corresponds to the observed flux and
  the upper limit to the maximum flux predicted by a synchrotron
  bubble model (see references 2 and 20 for details).  We adopt these
  limits to compensate for sparse radio coverage.}

\label{tab}
\end{deluxetable*}

\begin{figure*}
{\includegraphics[clip=true, angle=90,width=15.85cm]{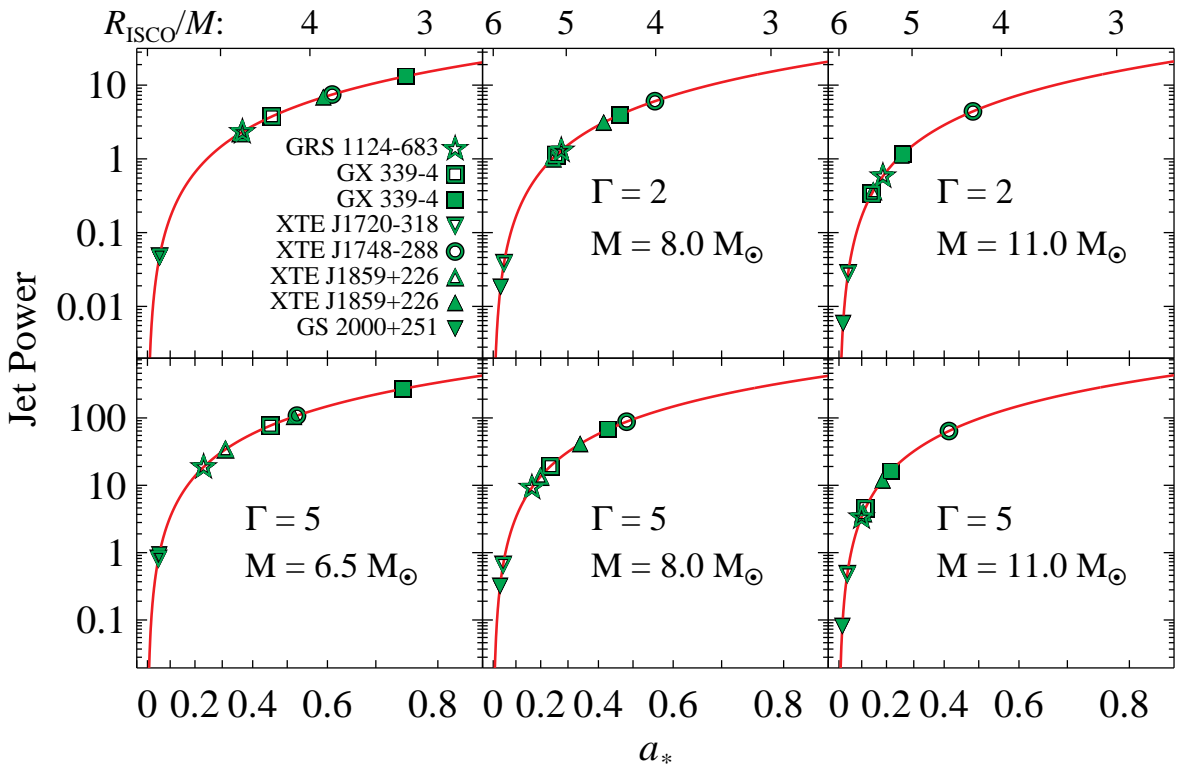}}
\caption{In successive panels, we plot the spin for each black hole
  listed in Table~\ref{tab} that results from adopting a value of $M$
  (6.5 $\msun$, 8$\msun$, and 11$\msun$, from left to right) and a
  value of $\Gamma$ (2 in the upper row and 5 in the lower).  The red
  line in each panel is a fit to the NM12 model using the data for all
  five calibration sources (Section~\ref{section:h1743}). In producing
  this figure, we fixed $M_2=0.3~\msun$ for the four systems with
  measured values of the mass function, and arbitrarily show results
  using $i=60\degr$ for the two without.}\label{fig:panel}
\end{figure*}

Our results are illustrated in Figure~\ref{fig:panel}.  Spin estimates
for each black hole are shown in the individual panels, which
correspond to our two bracketing values of $\Gamma$ (top and bottom
rows) and to three values of $M$ (increasing from left to right),
namely the median and 1$ \sigma$ limits of the \citealt{Ozel_2012}
distribution.  Every panel shows a pair of spin values for each of the
black holes that have two distance estimates (Table~\ref{tab}).


\section{Comparison with Fe-line Measurements}\label{section:feline}

The relationship between black hole spin and radio jet power (both
here and in NM12) has to this point been explored using spin data
obtained exclusively by applying the continuum-fitting method. We now
investigate pertinent black hole spin data obtained using the Fe-line
method. Here, black hole spin is inferred from the breadth and shape
of spectral fluorescence features, which are produced in the
strong-gravity environment of the inner accretion disk (e.g.,
\citealt{Fabian_1989}).

\begin{figure*}
{\includegraphics[clip=true, angle=90,width=15.85cm]{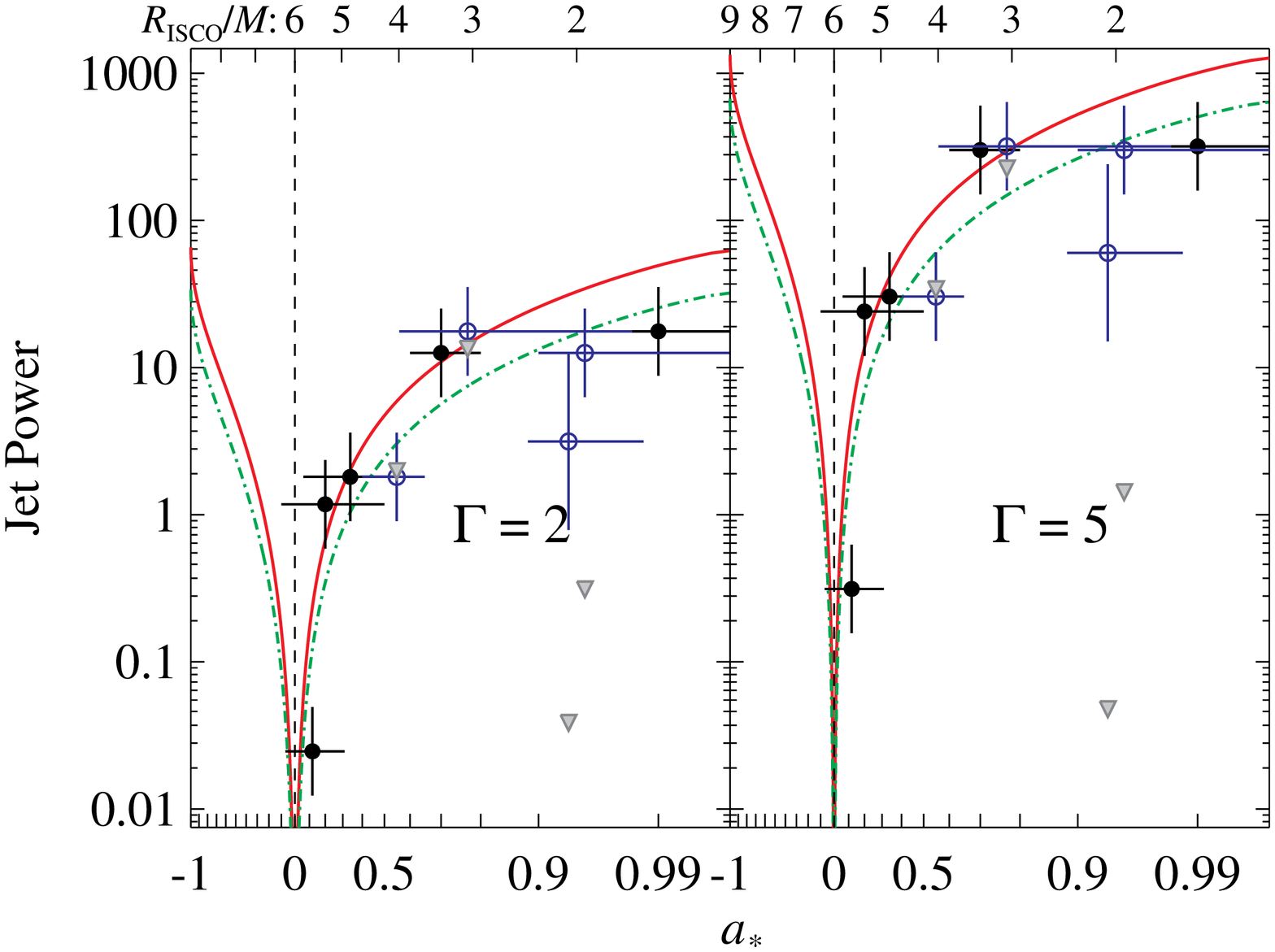}}
\caption{Fe-line spin measurements (open symbols) are compared against
  the best-fitting NM12 model -- the solid curve -- which is based on
  continuum-fitting spin measurements and a fit to the five
  calibrations sources (filled circles) shown both here and in
  Figure~\ref{fig:h1743}.  A dot-dashed curve shows the best fit
  achieved to the Fe-line and continuum-fitting measurements together.
  Both sets of fits have been computed by assuming that all jets are
  produced with either $\Gamma=2$ (left panel) or $\Gamma=5$ (right
  panel).  The jet powers obtained using the fitted inclinations
  returned by the Fe-line fits are plotted as filled gray
  triangles.}\label{fig:feline}
\end{figure*}

Among the black holes that we have so far considered, four have
Fe-line spin estimates: GRS 1915+105 ($\spin = 0.6-1$;
\citealt{Blum_2009}\footnote{Two possible solutions are reported:
$\spin = 0.56\pm0.02$ and $\spin = 0.98\pm0.01$.}), GRO J1655--40
($\spin > 0.9$; \citealt{Reis_2009}), GX 339--4 ($\spin \approx 0.93$;
\citealt{Miller_GX339, reisgx}), and XTE J1550--564 ($\spin \approx
0.55^{+0.10}_{-0.15}$; \citealt{Steiner_j1550spin_2011}).  The
reported errors for GX~339--4 and GRS~ 1915+105 are statistical and
small (0.01--0.02); in these cases, we adopt $\Delta\spin=0.05$ as a
rough estimate of the systematic uncertainty.

In considering the Fe-line spin data for these four sources, there are
two natural choices for the inclination of the black hole spin vector:
the axis perpendicular to the binary orbital plane, or the disk
inclination returned from the Fe-line spectral fits.  Throughout this
paper, we have adhered to the former by assuming that the black hole's
spin is aligned with the orbital angular momentum vector.  This
assumption is motivated and reasonable because the timescale for
alignment (e.g., \citealt{Martin_j1655_2008}) is an order of magnitude
smaller than the binary lifetime, so that nearly all of the several
dozen black holes in known transient systems should presently be well
aligned (e.g., see \citealt{Steiner_j1550jets}, and references
therein).

For the four sources in question, the Fe-line spectral fits return the
following estimates of disk inclination (which are determined largely
by the blue wing of the fluorescent line features): GRS 1915+105
($i = 55\degr - 70\degr$), GRO J1655--40 ($i = 30\degr^{+5}_{-10}$),
GX 339--4 ($i = 20\degr\pm1\degr$), XTE J1550--564 ($i = 71\degr -
82\degr$).  We caution that these inclination estimates are subject to
a systematic uncertainty of order $\sim10\degr$ for three reasons: The
spectral models employed in these fits (1) have not yet accounted for
radial variation of ionization across the face of the disk, which can
modify the structure of the blue wing of the line profile; (2) omit
treatment of Fe $K\beta$ and other (high-order) line transitions that
are most important at low or moderate ionization and which can
contribute flux just blueward of the dominant Fe \ka\ feature (more
recent Fe-line models have now incorporated many additional lines,
e.g., \citealt{Garcia_Kallman_2010}); and (3) provide only a cursory
treatment of the ``warm absorber'' features introduced by ionized disk
winds.  Warm absorbers, in the notable case of GRO~J1655--40, have
been found to vary with state and to contribute dozens of spectral
lines at high significance \citep{Miller_2006, Neilsen_2012}.

Although these three effects degrade estimates of inclination, which
depend on the extent of the blue wing of the line, they have a minor
affect on the spin parameter because it is determined principally by
the red wing of the line, which is an order of magnitude broader than
the energy shift induced in the blue wing by varying inclination (see
e.g., \citealt{Reis_2009, Reis_2012, Fabian_2012}).  Therefore, with a
reasonable degree of confidence, we make use of the fitted spins while
at the same time adopting the assumption of alignment, and so we
choose the orbital inclination angle over the inclination angle
returned by the Fe-line fits.
  
In Figure~\ref{fig:feline}, the Fe-line results are shown alongside
the data for our five calibration sources from Figure~\ref{fig:h1743}.
Because none of the Fe-line sources has a low spin, the Fe-line data
alone only weakly test the NM12 model.  At the same time, the Fe-line
and continuum-fitting results are reasonably consistent.  The
dash-dotted curve shows a best fit based on both the continuum-fitting
and Fe-line spin data.  The normalization for this fit is roughly a
factor of two lower than results from continuum-fitting alone
(primarily because of the relatively high Fe-line spin and low radio
flux from GX 339--4).  This shift is very small compared to the three
orders of magnitude spanned in jet power.

We note that our conclusion that the Fe-line spin measurements are
consistent with the NM12 model is contingent upon the use of orbital
inclinations.  That is, results obtained using the inclinations
returned by the Fe-line fits (shown by the solid triangle symbols in
Figure~\ref{fig:feline}) are inconsistent with the NM12 model.  This
is due to the large difference in the beaming correction implied by
the Fe-line inclinations for GX 339--4 and GRO~J1655--40.



\section{Discussion}\label{section:discussion}

An inspection of Figure~\ref{fig:panel} reveals that the inferred jet
power increases with $\Gamma$ because the inclination angles are
generally high ($i \gtrsim 35\degr$; see Section~\ref{section:model}
and Table~\ref{tab}) for the sample of six black holes in
Table~\ref{tab}.  We now consider the effect of increasing the mass
(for a fixed value of the mass function): This decreases $i$ and
increases the Doppler factor, thereby decreasing the inferred jet
power and the spin.  Thus, with the radio flux fixed at its observed
value, a high value of $M$ implies low spin.  In particular, for $M
\gtrsim 8\msun$, none of the black holes of Table~\ref{tab} is
expected to have a spin above $\spin \approx 0.5$.

Therefore, based on the NM12 model, the sample of six black holes in
Table~\ref{tab} is expected to have low masses, low spins, or both.
An additional outcome of the model is that, as an ensemble, the
transient black holes contrast starkly with the three wind-fed,
X-ray-persistent black holes, which have masses in the range $M =
11-16~\msun$ and spins ranging from $a_* = 0.85$ to $a_* > 0.95$
(\citealt{Gou_2011}, and references therein).  This suggests a
dichotomy between the black holes that form in these two distinct
classes of binary systems.

Table~\ref{tab} shows that each of the predicted black hole spin
values is expected to be below $\spin = 0.7$.  Furthermore, two
sources, XTE J1720--318 and GS 2000+25, may have record low values of
spin.  Measuring their spins directly will provide a strong test of
the NM12 model.  We note that our result for GX 339--4, $\spin < 0.6$,
is consistent with the upper limit from continuum-fitting by
\citet{Kolehmainen_Done_2010}.

In addition to predicting the spins of six black holes, we have also
provided estimates of the orbital inclination angles of their host
binaries by using the semiempirical mass distribution of
\citet{Ozel_2012}.  Future ground-based and X-ray studies will sharpen
up the estimates of $M$, $i$ and $D$ that we have used here and can
lead to direct measurements of spin, thereby testing the predictions
laid out in Table~\ref{tab}.

\section{Conclusions}\label{section:concs}

Under the assumption that the 5~GHz peak radio flux density from
transient black holes can be used as a proxy for their jet power, the
work of \citet{NM12} demonstrates a clear and empirical link between
jet power and black hole spin, as predicted by \citet{BZ77}.  The
addition of a fifth calibration source, H1743--322, whose spin and jet
power are entirely consistent with the NM12 model, strengthens that
link.  For a moderate and appropriate range of jet speeds
($\Gamma=2-5$), we have used the jet-power vs.\ spin correlation for
all five calibration sources to predict the spins of six black hole
primaries located in transient systems, which contain low-mass
secondaries.  Surprisingly, all of these predicted spins are
relatively low, $\spin < 0.7$, especially when compared to the spins
of the black hole primaries in the persistent and wind-fed systems
($\spin > 0.85$), which have massive companions.  Future measurements
of spin can be used to test these predictions and the NM12 model.

\acknowledgments 

J.F.S. was supported by the Smithsonian Institution Endowment Funds.
J.E.M. acknowledges support from NASA grant NNX11AD08G, and
R.N. acknowledges support from NASA grant NNX11AE16G.  We thank Robert
Dunn for input on his work, Rubens Reis and Tom Maccarone for
constructive discussions, and appreciate feedback on this manuscript
from Laura Brenneman and Andy Fabian.  We are grateful to Pawan Kumar 
for helpful discussions on jet emission.

\newcounter{BIBcounter}        
\refstepcounter{BIBcounter}

\appendix

\section{A Basis for the Standard Candle
  Assumption}\label{append:candle}

\setcounter{table}{0}
\renewcommand\thetable{\Alph{section}\arabic{table}}

In Section~\ref{section:Intro}, we assert that at X-ray maximum a
black hole transient approaches its Eddington limit and that it
therefore reasonably approximates a standard candle.  This is a
crucial assumption because it implies that the accretion power
($\dot{M}/M$) is roughly the same in different objects when they
exhibit ballistic jets.  This allows us to meaningfully compare the
jet powers (i.e., the peak mass-scaled radio luminosities) of the
various sources in our sample, which vary by a factor of 700 for a
uniform assumed value of $\Gamma=2$ and by a factor of 1000 for
$\Gamma=5$ (Fig.~\ref{fig:h1743}).

It is problematic to test this standard candle assumption for several
reasons.  The most important of these is that in nearly all
observations of black hole transients much of the X-ray flux falls
outside the passband of the detector, and, at the same time, there is
no standard model one can use to make a bolometric correction.  With
this in mind, we compute {\it firm lower limits} to the peak
luminosities of these sources using model fluxes reported in the
literature that fall within the passband of the detector in question.
This straightforward and empirical approach necessarily yields
underestimates of peak luminosity because it ignores flux at the high-
and low-energy ends of the spectrum.

In our discussion, we consider only black hole transients that have
made a hard-to-soft transition.  In particular, we disregard the
several systems discussed by \citet{Brocksopp_2004} that have never
made this transition, and we also disregard ``failed'' outbursts of
other systems that stalled in the low/hard state (e.g., the 2001 and
2002 outbursts of XTE J1550--564; see Fig.\ 6$a$ in \citealt{RM06}).
In Section~\ref{append:candle1} below, we consider only the largest
outburst that has been observed for each of the sources in our sample.
We show that each of our five calibration sources, which have
relatively high quality distance estimates, reaches about half or more
of its Eddington limit during a major outburst.  In
Section~\ref{append:candle2}, we further show that even if the
distance is poorly known one can conclude that the peak luminosity of
a black hole transient is at least $\approx10$\% of Eddington if the
source has made a hard-to-soft transition.

\subsection{Peak Luminosities during Major Outbursts}\label{append:candle1}

Our determinations of the peak observed component of the
Eddington-scaled luminosities of ten sources are given in the two
rightmost data columns in Table~\ref{tab:appendix}: $L_{\rm
Peak}/L_{\rm Edd}$ is the isotropic (Eddington-scaled) luminosity and
$L_{\rm Disk, Peak}/L_{\rm Edd}$ is the luminosity assuming that the
emitter is a thin disk (see Table footnote $b$ in
Table~\ref{tab:appendix}).  In the former case, the mean observed
component of luminosity for the the five calibration sources in the
top half of the table is $0.57\pm0.25$ (std.\ dev.), and in the latter
case it is $0.76\pm0.30$ (where we have assumed that GRS 1915+105 is
at its Eddington limit).  {\it Thus, we conclude that our calibration
sources typically reach half or more of their Eddington limit during a
major outburst}.

We note that our conclusion is corroborated by an independent analysis
for a sample of black hole transients considered by \citet{Dunn_2010}.
Their results for the peak luminosities of these sources, all of which
have undergone a hard-to-soft transition, are summarized in their
Figure~11.  For the four recurrent sources (XTE~J1550--564,
4U~1630--47, GX~339--4 and H1743--322), we restrict our attention to
the brightest outburst for each source.  For H1743 only, we correct
the luminosity given by Dunn et al.\ using $D=8.5$~kpc
\citep{Steiner_2012_H1743} in place of their guess of 5 kpc.  We
disregard SLX 1746-331 whose distance is essentially unconstrained
within the Galaxy.  The mean isotropic luminosity of the remaining
sample of eight sources is $L_{\rm Peak}/L_{\rm Edd} = 0.43 \pm 0.23$
(std.\ dev.), which is quite comparable to our result quoted above.
Dunn et al.\ obtain a somewhat lower value than we do because they
consider only the 2--10 keV component of luminosity while we generally
consider wider bandpasses (see below).

We now present the details of our analysis that support the italicized
conclusion stated above.  The sources listed in the top half of
Table~\ref{tab:appendix} are our calibration sources
(Fig.~\ref{fig:h1743}), and those in the lower half are the sources
listed in Table~\ref{tab}. (We disregard XTE J1720--318 because there
are no suitable X-ray data near maximum).  The distance and mass
estimates are taken from Table 1 in NM12 and Table~\ref{tab} herein,
except for the six black holes that lack mass measurements; for these
we adopt the nominal value $M=8~M_{\odot}$
\citep{Ozel_2010,Farr_2011}.  As discussed below, for all but two
sources, GRO J1655--40 and GRS 1915+105, the tabulated value of
$L_{\rm Peak}/L_{Edd}$ at X-ray maximum was computed for the
corresponding radio outburst used in estimating the jet power.  In all
cases, our peak luminosities are based on the peak unabsorbed fluxes
that have been reported in the literature for the missions and
bandpasses listed in Table~\ref{tab:appendix}.

The firm lower limits on peak luminosities given in
Table~\ref{tab:appendix} are strictly empirical; i.e., they are
computed directly from the observed maximum fluxes $F_{\rm max}$
assuming either an isotropic source, $L_{\rm Peak}/L_{Edd} =
{{4\pi}{D^2}{F_{\rm max}}}/({{1.3}{\times}{10^{38}}{M/M_{\odot}}}$),
or alternatively a thin disk (see footnote $b$ in
Table~\ref{tab:appendix}).

{\it A0620--00:} \citet{Doxsey_1976} report a 1--10 keV flux at the
maximum of the 1975 outburst of $F_{\rm max} (1-10~$keV$) = 1.7 \times
10^{-6}$ for a thermal bremsstrahlung spectrum with $kT=1.7$ keV.
Sixteen days later, they made a second observation, this time
additionally employing the SAS--3 low-energy system (0.15--0.9 keV).
The spectral parameters and flux derived for this latter observation
allow one to conclude that the 0.3--1 keV flux was 2.04 times the
1--10 keV flux.  Taking this result as a guide and using the spectrum
determined by \citeauthor{Doxsey_1976} at maximum, we find that the
0.3--1 keV flux at that time was 1.77 times the 1--10 keV flux.  We
therefore conclude: $F_{\rm max} (0.3-10~$keV$) = 3.0 \times
10^{-6}$~ergs~cm$^{-2}$~s$^{-1}$..

{\it XTE J1550--564:} Here we use the X-ray flux reported at the peak
of the extraordinary 7-Crab flare, which was observed on 1998
September 19, and which preceded the detection of the radio ejection
by four days \citep{Hannikainen_2009}.  We adopt the 2--20 keV and
20--100 keV fluxes reported by \citet{Sobczak_2000} respectively in
their Tables 3 \& 4: $F_{\rm max} (2-100~$keV$) = 2.72 \times
10^{-7}$~ergs~cm$^{-2}$~s$^{-1}$.

{\it GRO J1655--40:} Major X-ray outbursts of this source were
observed in 1994, 1996 and 2005.  Unfortunately, radio data at X-ray
maximum were obtained only for the 1994 outburst (see Table~1 in
NM12), and the available X-ray data at maximum for this outburst
(BATSE at $E>20$ keV) do not provide a useful lower limit on $L_{\rm
  Peak}/L_{Edd}$.  Therefore, in estimating $L_{\rm Peak}/L_{Edd}$ for
this source, we consider the well-observed 1996 and 2005 outbursts,
which had very comparable peak intensities of $\approx 300~{\it
  RXTE}$~ASM counts~s$^{-1}$, 2--12 keV \citep{Sobczak_j1655_1999,
  Brocksopp_2006}.  As our proxy for the 1994 X-ray peak flux,
we adopt the peak flux observed for the 2005 outburst on May 16 by
\citeauthor{Brocksopp_2006} because the {\it Swift} XRT and BAT
detectors provide broadband coverage with superior low-energy
coverage; we find: $F_{\rm max} (0.7-150~$keV$) = 2.26 \times
10^{-7}$~ergs~cm$^{-2}$~s$^{-1}$.

{\it H1743--322:} As in the case of XTE J1550--564, and as discussed
in Section~\ref{section:h1743}, the jet was launched by an impulsive
power-law flare, which was observed on 2003 May 6 (MJD 52765.9); the
radio flux reached a maximum $\approx2.6$ days later
\citep{JEM_H1743}.  The peak X-ray flux reported in Table~A2 of
\citeauthor{JEM_H1743} is $F_{\rm max} (2-100~$keV$) = 5.84 \times
10^{-8}$~ergs~cm$^{-2}$~s$^{-1}$.

{\it GS 1915+105:} As in the case of GRO 1655--40, poor X-ray coverage
does not allow us to set a useful lower limit on $L_{\rm
  Peak}/L_{Edd}$ for either of the well-studied radio outbursts
considered in NM12 \citep{Rodriguez_1995, Fender_1999}.  Furthermore,
the distance to this source is quite uncertain, ranging from about 7
kpc to above 12 kpc (see Fig.\ 18 in \citealt{McClintock_2006}).  We
therefore fall back on the widely accepted conclusion that this source
is generally exceptionally luminous.  For example, \citet{Done_2004}
infer luminosities as high as $\approx 1.7$ Eddington for $D=12.5$ kpc
(or 1.0 Eddington for 9.5 kpc).  We therefore assume, as indicated in
Table~\ref{tab:appendix}, that the source was near its Eddington limit
at the time of peak radio emission on 1994 March 24
\citep{Rodriguez_1995}.

{\it GRS 1124--683 (Nova Mus 1991):} In their Table 2,
\citet{Ebisawa_1994} summarize the spectral parameters and fluxes for
frequent observations of this source during its 1991 outburst.  For
the peak-flux observation of 1991 January 16 at 19 hours UT, we adopt
their tabulated value of the hard flux in the 2--20 keV band and,
using the spectral parameters given for the thermal component, compute
the corresponding soft flux in this same band and conclude: $F_{\rm
  max} (2-20~$keV$) = 1.53 \times 10^{-7}$~ergs~cm$^{-2}$~s$^{-1}$.

{\it GX 339--4:} The times of the peak radio flux \citep{Gallo_2004}
and the peak {\it RXTE} ASM count rate coincided within roughly one day.
We adopt the peak X-ray flux plotted in Figure 9 (panel b) in
\citet{RM06}: $F_{\rm max} (2-20~$keV$) = 2.20 \times
10^{-8}$~ergs~cm$^{-2}$~s$^{-1}$.

{\it XTE J1748--288:} Both the X-ray and radio coverage are relatively
spotty for this source \citep{Brocksopp_2007}.  In estimating the peak
X-ray flux, we use the results of the analysis of an {\it RXTE} PCA
spectrum that was obtained at the time of peak intensity as recorded
by the {\it RXTE} ASM.  This spectrum is plotted in Figure 4.14 and
the spectral parameters are tabulated in Table 4.4 in \citet{MR06}.
Using these data, we find: $F_{\rm max} (2-20~$keV$) = 3.06 \times
10^{-8}$~ergs~cm$^{-2}$~s$^{-1}$.

{\it XTE J1859+226:} As in the case of GX 339--4, the radio flux and
the {\it RXTE} ASM count rate peaked within about a day of each other.
We adopt the peak X-ray flux plotted in Figure 8 (panel b) in
\citet{RM06}: $F_{\rm max} (2-20~$keV$) = 3.55 \times
10^{-8}$~ergs~cm$^{-2}$~s$^{-1}$.

{\it GS 2000+25:} The peak of the outburst occurred on 1988 April 28
\citep{Tsunemi_1989}.  As a lower bound, we adopt the flux for an
observation made two days after the peak, on April 30, which is
reported by \citet{Terada_2002} in their Tables 2--5: $F_{\rm max}
(1.7-37~$keV$) = 2.6 \times 10^{-7}$~ergs~cm$^{-2}$~s$^{-1}$.

{\it XTE J1720--318:} While we give a spin estimate for this source
(Table~\ref{tab}), we exclude it in Table~\ref{tab:appendix} because
(apart from the {\it RXTE} ASM) there is no information on its
spectrum at maximum.  If we make the arbitrary assumption that its
spectrum and flux is the same as that of XTE J1748--288 (see above),
then adopting $M=8~M_{\odot}$ and the very uncertain distance estimate
of $D=6.5$~kpc, we find $L_{\rm Peak}/L_{\rm Edd} = 0.15$.  (We note
that for this source, even at twice the nominal distance, the spin
prediction remains very low, $\spin \lesssim 0.14$.)

\subsection{A Floor on the Peak Luminosity}\label{append:candle2}

Finally, we present evidence for a floor on the peak luminosity of
systems that undergo a hard to soft transition through the thermal
state.  As mentioned at the outset of Section 4, such a state
transition is one of our selection criteria.  Setting a minimum peak
luminosity is important for sources whose distances are relatively
uncertain, such as most of the sources listed in Table~1.  We again
use the luminosity data summarized by \citet{Dunn_2010} in their
Figure~11 (and we again exclude SLX 1746--331 and adopt $D=8.5$~kpc
for H1743--32).

For each of the four recurrent sources, Dunn et al. plot the peak
luminosities for between two and four separate outburst cycles.
Considering now the faintest outburst for each of the four recurrent
sources, we conclude that the peak luminosity of all eight sources in
the Dunn et al.\ sample exceeds 8\% of Eddington, which is to be
compared to the factor of $\sim1000$ range in the peak radio
luminosities of our five calibration sources.  Again, this floor of
8\% of Eddington is a very conservative lower limit because Dunn et
al.\ consider only the 2--10 keV component of luminosity.

   \begin{deluxetable}{cccccccc}
\small
  \tabletypesize{\scriptsize} 
  \tablewidth{0pt}  
  \tablecaption{Observed Component of Luminosity at Outburst Maximum}
  \tablehead{\colhead{Object} & \colhead{$D$(kpc)} & \colhead{$M (M_{\odot}$)} & \colhead{Mission} & \colhead{Band (keV)} & \colhead{$L_{\rm Peak}/L_{\rm Edd}$\tablenotemark{a}} & \colhead{$L_{\rm Disk, Peak}/L_{\rm Edd}$\tablenotemark{b}} & \colhead{References}}
  \startdata
  A0620-00       &   1.06  &   6.6     &  SAS-3       & 0.3--10   &   0.47                  & 0.37     & 1        \\
  XTE J1550--564 &   4.38  &   9.1     &  RXTE        & 2--100    &   0.53                  & 1.00    & 2        \\
  GRO J1655--40  &   3.2   &   6.3     &  Swift       & 0.7-150   &   0.34\tablenotemark{c} & 0.50\tablenotemark{c}     & 3        \\
H1743--322     &   8.5   &   8       &  RXTE        & 2--100    &   0.49                    & 0.94     & 4        \\
GRS 1915+105   &   11    &   14      &  RXTE        & 2--20    &$\sim1$\tablenotemark{c}   & $\sim1$\tablenotemark{c}  & 5,6      \\    
\hline
GRS 1124--683  &   5.9   &   8       &  Ginga       & 2--20     &   0.61                    & 0.48     & 7        \\
GX 339--4      &   8     &   8       &  RXTE        & 2--20     &   0.16                    & 0.20     & 8        \\
               &   15    &           &              &           &   0.57                     & 0.69     &          \\ 
XTE J1748--288 &   8     &   8       &  RXTE        & 2--20     &   0.23                    & 0.16\tablenotemark{d}     & 6        \\
XTE J1859+226  &   8     &   8       &  RXTE        & 2--20     &   0.26                    & 0.26     & 8        \\        
               &   14    &           &              &           &   0.80                     & 0.80     &          \\
GS2000+25      &   2.7   &   8       &  Ginga       & 1.7--37   &   0.22                    & 0.24     & 9        \\
\enddata

\tablerefs{(1) \citealt{Doxsey_1976}; (2) \citealt{Sobczak_2000}; (3) \citealt{Brocksopp_2006}; 
(4) \citealt{JEM_H1743}; (5) \citealt{Done_2004}; (6) \citealt{MR06}; 
(7) \citealt{Ebisawa_1994}; (8) \citealt{RM06}; (9) \citealt{Terada_2002}.}

\tablenotetext{a}{The observed peak luminosity in the passband
  indicated, assuming isotropic emission.}

\tablenotetext{b}{The observed peak luminosity assuming thin disk geometry,
  i.e., accounting for the inclination according to $L_{\rm Disk,
    Peak} = (L_{\rm Peak}/L_{\rm Edd}) / 2{\rm cos}~i$.  Inclinations are from
  Table~1 and, for the original calibration sources, from Table~1
  of \citet{NM12}}.

\tablenotetext{c}{In these two cases only, the luminosity was computed for an X-ray 
outburst different than the one used in estimating the jet power; see text.}

\tablenotetext{d}{For $i=45\degr$.}

\label{tab:appendix}
\end{deluxetable}

\section{Estimating the Energy of Ballistic Synchrotron Bubbles}\label{append:blob}

In the toy analysis that follows, we derive a relationship between
synchrotron emission from a plasmoid and its bulk kinetic energy.
This derivation is not intended to be rigorous.  Rather, our aim is to
demonstrate that a roughly linear relationship between synchrotron
flux density at light-curve maximum and blob kinetic energy -- as
assumed by the empirical NM12 model -- is a natural outcome of
classical jet theory.  We stress that this derivation is applicable to
impulsive, ballistic jets (e.g., \citealt{Mirabel_Rodriguez_1994,
Hjellming_Rupen}) as opposed to steady-state jets, which are described
by a different class of models (e.g., \citealt{Heinz_Sunyaev_2003,
Fender_2001, Falcke_1996}).  For additional background on the
synchrotron-bubble model discussed here, we refer the interested
reader to \citet{van_der_Laan_1966, Kellerman_Owen_1988,
Hjellming_1988, Hjellming_Johnston}.

We assume that all beaming-related effects ($\Gamma$ dependence) have
been removed. That is, we work in the frame of a single radiating
blob.  We are interested in the relation between the radio luminosity
of the blob at 5\,GHz and the energy of the blob.  Using a fairly
standard set of assumptions for synchrotron-emitting blobs (e.g.,
\citealt{Hjellming_Johnston}), we derive such a relationship.  The
following calculation is approximate and ignores certain factors of
order unity, but it is dimensionally correct.

Let $B$ be the magnetic field strength, and let $\gamma$ be the
typical Lorentz factor of the electrons that produce synchrotron
radiation at 5\,GHz.  From standard synchrotron theory,
\begin{equation}
\nu_{\rm synch} = \frac{3}{4\pi}\,\gamma^2\, \frac{eB}{m_e c} = 5\,{\rm GHz},
\label{eq:nusynch}
\end{equation}
which gives
\begin{equation}
B = 1200\, \gamma^{-2} ~{\rm G}.
\label{eq:B}
\end{equation}
Let us assume that the electrons in the blob have an energy distribution
of the form
\begin{equation}
N(\gamma)\,d\gamma = N_0\, \gamma^{-p}\, d\gamma,
\quad \gamma \geq 1,
\label{eq:dNdgamma}
\end{equation}
where for simplicity we assume that the minimum energy of the
electrons is $\gamma_{\rm min}\approx 1$. The effective number of
electrons radiating at 5\,GHz is then
\begin{equation}
N_\gamma \approx \gamma\,N(\gamma) = N_0\, \gamma^{-(p-1)}.
\label{eq:Ngamma}
\end{equation}
The synchrotron luminosity of these electrons is
\begin{equation}
\nu L_\nu \approx N_\gamma\, \frac{2e^4}{3m_e^2c^3}\,
\gamma^2 B^2\, \frac{1}{2} = 1.1\times10^{-9} \,N_\gamma \gamma^{-2} ~
{\rm erg\,s^{-1}},
\label{eq:nuLnu}
\end{equation}
where the factor of $1/2$ in the middle expression is to allow for the
fact that $d\ln\nu = 2\,d\ln\gamma$ (eq.~\ref{eq:nusynch}).

Let us assume that there is rough equipartition between the energy in
the magnetic field and that in the relativistic electrons. Assuming
that the radius of the blob is $R$, we write the equipartition
condition as
\begin{equation}
\frac{B^2}{8\pi}\,\frac{4\pi}{3}R^3 = \xi\, N_0\,m_e c^2,
\label{eq:equi}
\end{equation}
where the dimensionless number $\xi$ measures the deviation from
strict equipartition.

We now make use of the fact that, at the peak of the radio
light-curve, the synchrotron radiation makes a transition from
self-absorbed radiation to optically thin emission. Writing the
effective temperature $T_\gamma$ of the relevant electrons as
\begin{equation}
kT_\gamma = \gamma m_e c^2,
\label{eq:Tgamma}
\end{equation}
the condition of marginal self-absorption requires the radio
luminosity at the light-curve maximum to satisfy
\begin{equation}
\left(\nu L_\nu\right)_{\rm max} = 1.1\times10^{-9} \,N_\gamma
\gamma^{-2} ~ {\rm erg\,s^{-1}} = 4\pi R^2\,\pi\nu\,
2\,\frac{\nu^2}{c^2}\, kT_\gamma.
\label{eq:selfabs}
\end{equation}
We have used the Rayleigh-Jeans approximation for the expression on
the right.

For ease of comparison with our previous work and with the discussion
in the main text, we express the luminosity $(\nu L_\nu)_{\rm max}$ in
terms of the quantity $P_{\rm jet}$ (see eq.~\ref{eq:pjet}):
\begin{equation}
P_{\rm jet} \equiv \frac{(\nu S_\nu)_{\rm max} D^2}{M} ~{\rm kpc^2\,GHz\,Jy\,M_\odot^{-1}},
\end{equation}
which is defined in practical units. In what follows, we assume for
simplicity that $M=10M_\odot$. Hence
\begin{equation}
\left(\nu L_\nu\right)_{\rm max} = 4\pi\, \left(\nu S_\nu\right)_{\rm
  max}\, D^2 = 1.2\times10^{31} P_{\rm jet} ~{\rm erg\,s^{-1}}.
\end{equation}

We have collected enough relations to solve for all quantities in
terms of the single observable quantity $P_{\rm jet}$. Let us assume
that $p=5/2$, which corresponds to an optically thin synchrotron
spectrum $S_\nu \propto \nu^{-0.75}$. For this reasonable value of
$p$, we obtain
\begin{eqnarray}
B &\approx& 2.4\, \xi^{2/9} P_{\rm jet}^{-1/9} ~{\rm G}, \\
\gamma &\approx& 22\, \xi^{-1/9} P_{\rm jet}^{1/18}, \\
N_0 &\approx& 5.5\times 10^{44}\, \xi^{-7/18} P_{\rm jet}^{43/36}, \\
R &\approx& 7.7\times 10^{12}\, \xi^{1/18} P_{\rm jet}^{17/36} ~{\rm cm}.
\end{eqnarray}
In obtaining these results, we have assumed that $P_{\rm jet}$ is
measured from the peak radio luminosity at 5\,GHz (in fact, all our
relations assume $\nu=5$\,GHz). Interestingly, the equipartition
factor $\xi$ turns out to be relatively unimportant.

The above results allow us to estimate various quantities in the frame
of the blob. To calculate the relativistic bulk kinetic energy of the
blob in the ``lab'' frame, we assume that there is one proton for each
electron, i.e., a total of $\approx N_0$ protons in the blob. Since
the thermal energy in the electrons is small compared to the rest mass
energy of the protons, we expect the protons to be effectively cold.
Let the blob move with bulk Lorentz factor $\Gamma$ in the lab frame.
In this frame, the blob energy is dominated by the proton kinetic
energy. Hence, we estimate the energy in the blob to be
\begin{equation}
E_{\rm blob} \approx N_0\, \Gamma m_pc^2 \approx 8.3\times10^{41}
\,\Gamma\,\xi^{-7/18} P_{\rm jet}^{43/36} ~{\rm erg}.
\label{eq:Eblob}
\end{equation}

The numerical values we have obtained above should not be taken too
seriously considering the approximations we have made. However, they
are reasonable.  For instance, for the microquasar GRS~1915+105, if we
take $P_{\rm jet} \sim 100$, we estimate the rest mass of the blob to
be $N_0 m_p \approx 2\times 10^{23}\,{\rm g}$, and for $\Gamma\sim5$
we find the bulk kinetic energy to be $\sim 10^{45} \,{\rm
  erg}$. These estimates are fairly close to those obtained by
\citet{Rodriguez_1999} even though they followed a different approach,
using the angular size of the blob instead of the light-curve maximum.

For our present purposes, the key result from the above analysis is
the scaling in equation (\ref{eq:Eblob}), which shows that the bulk
kinetic energy of the blob is expected to vary approximately as the
1.2 power of the 5~GHz radio power ($P_{\rm jet}$) at light-curve
maximum, i.e., there is a more or less linear relation between the two
quantities.  This provides strong support for our reliance upon
$P_{\rm jet}$ as a measure of the jet kinetic power.

\end{document}